\documentstyle[11pt]{article}

\input{psfig}

\catcode`\@=11 \def\@cite#1{[#1]}                  % \uppercite
\def\@cite#1{\raisebox{1ex}{{\footnotesize #1}}} % Produces the output of \cite
\catcode`\@=12                                     % \uppercite
\begin{document}
\begin{titlepage}

\begin{flushright}
SCCS-526 \hspace*{1.0cm} \\[2.5em]
\end{flushright}

\baselineskip=18pt

\begin{center}
{\LARGE
Analysis of Random Number Generators\\
Using Monte Carlo Simulation \\}
\bigskip
\bigskip
{\large
P.~D.~Coddington\\}
\smallskip
{\it
Northeast Parallel Architectures Center, Syracuse University,\\
111 College Place, Syracuse, NY 13244, U.S.A.\\}
\bigskip
\bigskip
September 14, 1993 \\
\end{center}
\bigskip
\bigskip

\begin{abstract}
\noindent
Monte Carlo simulation is one of the main applications involving
the use of random number generators. It is also one of the best
methods of testing the randomness properties of such generators,
by comparing results of simulations using different generators
with each other, or with analytic results.
Here we compare the performance of some popular random number
generators by high precision Monte Carlo simulation of the 2-$d$ Ising
model, for which exact results are known, using the Metropolis,
Swendsen-Wang, and Wolff Monte Carlo algorithms.
Many widely used generators that perform well in standard statistical
tests are shown to fail these Monte Carlo tests.
\end{abstract}

\end{titlepage}
% _____________________________________________________________________

%\begin{document}
\baselineskip=18pt

%%%%%%%%%%%%%%%%%%%%%%%%%%%%%%%%%%%%%%%%%%%%%%%%%%%%%%%%%%%%%%%%%%%
\section{Introduction}
%%%%%%%%%%%%%%%%%%%%%%%%%%%%%%%%%%%%%%%%%%%%%%%%%%%%%%%%%%%%%%%%%%%
\noindent
Monte Carlo simulation is an important numerical technique for studying
a wide range of problems in the physical sciences.\cite{MONTECARLO}
Being a probabilistic technique, it relies heavily on the use of pseudo-random
number generators.\cite{KNUTH,JAMES,LECUREV}
The generation of random numbers on a computer is a notoriously difficult
problem.
An ideal random number generator would provide numbers that are
uniformly distributed,
uncorrelated,
satisfy any statistical test of randomness,
have a large period of repetition,
can be changed by adjusting an initial ``seed'' value,
are repeatable, portable,
and can be generated rapidly using minimal computer memory.

Many statistical tests have been developed to
check for randomness,\cite{KNUTH,MARSAGLIA} and in most cases the
period of the generator can be calculated (at least approximately).
As noted in a number of recent review articles,\cite{JAMES,LECUREV,PARK}
random number generators provided by computer vendors or recommended
in computer science texts often have been (and unfortunately continue to be)
of poor quality.
Even generators that perform well in standard statistical tests for
randomness may be unreliable in certain applications, as has been
found in some Monte Carlo
simulations.\cite{KALLE,SANTABARB,FILK,MILCHEV,LANDAU,GRASS1,GRASS2}

There are two main types of random number generators for producing sequences
of pseudo-random integers {$X_i$}:
\begin{enumerate}
\item Linear congruential generators (LCGs)\cite{KNUTH,PARK}
      \begin{description}
      \item[\qquad\qquad] $X_{i} = A * X_{i-1} + B \quad \bmod M$
      \end{description}
      which we will denote by L($A,B,M$). The period is $M$ for suitably
      chosen $A$ and $B$ ($M-1$ if $B=0$).
\item Lagged Fibonacci generators (LFGs)\cite{KNUTH,MARSAGLIA}
      \begin{description}
      \item[\qquad\qquad] $X_{i} = X_{i-P} \odot X_{i-Q}$
      \end{description}
      which we will denote by F($P,Q,\odot$), $P > Q$,
      where $\odot$ is any binary arithmetic operation, such as
      $+, -, *$ or $\oplus$ (the bitwise exclusive OR function XOR).
      The arithmetic operations are done modulo any large integer value,
      or modulo 1 if the $X$'s are represented as floating point numbers
      in the interval [0,1), as can be done if the operation is $+$ or $-$.
      Multiplication is done on the set of odd integers.
      For $b$-bit precision $X$'s, the period is
      $(2^{P}-1)2^{b-1}$, or $(2^{P}-1)2^{b-3}$ for multiplication,
      for suitably chosen lags.\cite{MARSAGLIA}
\end{enumerate}

It is possible to find sets of parameters ($A,B,M$) or ($P,Q,\odot$)
for which these two types of generators work well for most
practical purposes, and it is possible to improve the performance of
these generators by increasing $M$ or $P$.\cite{MARSAGLIA}
There are practical limits on these two parameters: $M$ should not be
very much greater than machine precision
to avoid using slow multi-precision arithmetic,
and a large lag $P$ means storing a large array of previous numbers in
the sequence (the ``lag table'') which may be subject to memory constraints.
However on most modern computers adequate values of $M$ and $P$ can be found
which are well within these limits.

Linear congruential generators have two major defects.
The first is that the least significant bits of the numbers produced are
highly correlated, and a resultant ``scatter-plot'' of ordered pairs of
random floating point numbers in the interval [0,1) shows regular lattice
structure.\cite{MARSAGLIA,PLANES,LECUYER,NIEDERREITER}
They are also known to have long-range correlations, especially for intervals
which are a power of 2.\cite{KALLE,FILK,PERCUS,GROTHE}
Another problem is that for 32-bit integers the period of these generators
is at most $2^{32}$, or of order $10^9$.
On a modern RISC workstation capable of around
$10^8$ floating point operations per second, this period can be exhausted
in a matter of minutes. This can be alleviated by the use of 64-bit precision,
however the correlation problems still remain (although to a lesser degree).
In spite of these problems, LCGs with well-chosen parameters perform well in
most standard statistical tests, and an LCG
(unfortunately not always with well-chosen parameters!)
is provided as the default generator on many computer systems.

Lagged Fibonacci generators using arithmetic operations ($+,-,*$)
give good results in standard statistical tests with very modest lags on
the order of tens.\cite{MARSAGLIA}
When the binary operation used is XOR, these generators are referred
to as generalized feedback shift register generators.\cite{TAUSWORTHE,STOLL}
Marsaglia has shown that XOR is one of the worst operations one can
use in a generator of this type, and strongly recommends the use of
standard arithmetic operations that have much longer
periods and perform much better in statistical tests.\cite{MARSAGLIA}
Although shift register generators pass statistical tests when the lag is
large enough (of order hundreds),\cite{MARSAGLIA,TWC,FINNS}
very little (apart from the period) is known theoretically about these
generators, and
they have produced biased results in Monte Carlo studies of the Ising model
in two\cite{LANDAU} and three\cite{SANTABARB} dimensions,
and of self-avoiding random walks.\cite{GRASS1,GRASS2}

Mixing two different generators is believed to improve performance in some
cases,\cite{MARSAGLIA,LECUYER} and many
generators that perform well in statistical tests are of this kind.
Marsaglia has suggested a fast, simple Weyl (or arithmetic sequence)
generator\cite{SWC,RANMAR}
\begin{description}
\item[\qquad\qquad] $X_{i} = X_{i-1} - K   \quad \bmod M$,
\end{description}
with $K$ a constant relatively prime to $M$,
that can be effectively combined with a lagged Fibonacci generator.
Adding a Weyl generator also increases the period of the combined
generator by a factor of $M$ (the period of the Weyl generator).
L'Ecuyer\cite{LECUYER} has shown how to combine two different 32-bit LCGs
to produce a mixed generator that passes the scatter-plot test
and has a long period of around $10^{18}$, thus overcoming some of the
drawbacks of standard LCGs.
Although these mixed generators perform well in empirical tests, there
is little theoretical understanding of their behavior, and it is
quite possible that mixing two generators may introduce new defects of
which we are unaware. A good single generator may therefore be
preferable to a mixed generator.

LCGs have the advantage that we have a relatively good (although still
limited) theoretically understanding of their randomness properties.
They are known to be defective, but their defects are fairly well understood
(for example, the lattice structure of an LCG can be determined analytically
using the spectral test\cite{KNUTH}), and in practice they work quite well.
There is clearly a need for better random number generators, and LFGs and
mixed generators are prime candidates. However currently there is little or
no theoretical understanding of these and other generators, and they are
used mainly on the basis of their performance in statistical tests.
They are believed to overcome some of the flaws of LCGs, although this has
not been proven and they may possess other flaws of which we are unaware.
It is therefore extremely important to subject random number generators to a
wide variety of precise statistical tests.

%%%%%%%%%%%%%%%%%%%%%%%%%%%%%%%%%%%%%%%%%%%%%%%%%%%%%%%%%%%%%%%%%%%
\section{Monte Carlo Tests}
%%%%%%%%%%%%%%%%%%%%%%%%%%%%%%%%%%%%%%%%%%%%%%%%%%%%%%%%%%%%%%%%%%%
\noindent
One practical way to test a random number generator is to use it for
Monte Carlo simulation of the two dimensional Ising model.\cite{MONTECARLO}
This simple model has been solved exactly for a finite lattice,\cite{FISHER}
so that values of the energy and the specific heat (the variance of the
energy) of the system calculated from the Monte Carlo simulation can be
compared with the known exact values.

A number of different Monte Carlo algorithms can be used to simulate the
Ising model.  Here we will concentrate on the three most widely used methods:
the Metropolis algorithm,\cite{MONTECARLO,METROPOLIS}
which updates a single site of the lattice at a time;
the Swendsen-Wang algorithm,\cite{SW} which
forms clusters of sites to be updated collectively; and the Wolff
algorithm,\cite{WOLFF} which updates a single cluster of sites.
Each of these algorithms uses random numbers in a very different way.
The Swendsen-Wang and Wolff cluster update algorithms are extremely
efficient and allow very precise Monte Carlo simulations of the Ising model,
easily reducing statistical errors in the energy to better than one part
in $10^5$.
This precision provides us with a very effective practical test of the
randomness of a pseudo-random number generator, and in particular its
suitability for Monte Carlo simulation.

Ferrenberg {\it et al.}\cite{LANDAU} recently showed that some ``good''
random number generators, which perform well in standard statistical tests,
fail the ``Monte Carlo test''; that is, they produce incorrect results when
used in Monte Carlo simulations of the Ising model, especially using the
Wolff algorithm.
The generators studied by Ferrenberg {\it et al.} were:
\newcounter{bean}
\begin{list}{\roman{bean}.}{\usecounter{bean} \itemindent=10pt}
\item  CONG, the linear congruential generator
       L($16807,0,2^{31}-1$).\cite{KNUTH,PARK}
\item  Two shift register generators, F($250,103,\oplus$)
       and F($1279,1063,\oplus$).\cite{STOLL}
\item  SWC, a subtract-with-carry generator based on F($43,22,-$).\cite{SWC}
\item  SWCW, a combined subtract-with-carry and Weyl generator.\cite{SWC}
\end{list}
In spite of the premise of that paper, CONG and the shift register generators
are in fact known to be {\it not} good random number generators.
CONG has been recommended by a number of authors\cite{KNUTH,PARK}
as one of the best 32-bit linear congruential generators, however it still
suffers the small period and correlated low order bits of these generators.
Shift register generators have been criticized by Marsaglia, who showed that
those with small lags (less than 100) performed
poorly in statistical tests.\cite{MARSAGLIA} However similar
tests of F($250,103,\oplus$) gave good results,\cite{TWC,FINNS}
and Kirkpatrick and Stoll also obtained reasonable results with Monte Carlo
tests.\cite{STOLL}

Subtract-with-carry generators are another variation of LFGs, where
the standard operation of subtraction is replaced by subtraction with a
carry bit $C$, as follows:
\begin{description}
\item[\qquad\qquad] $X_{i} = X_{i-P} - X_{i-Q} - C,$
\item[\qquad\qquad] if $X_{i} \ge 0,
                       \quad C = 0,$
\item[\qquad\qquad] if $X_{i} < 0,
                       \quad X_{i} = X_{i} + M, \quad C = 1.$
\end{description}
This greatly increases the period of the LFG, to $M^P - M^Q$ for suitably
chosen $P,Q$ and $M$,\cite{SWC} compared to approximately $M 2^{P}$ for a
comparable LFG using subtraction.
We have used $M = 2^{32}-5$, which gives very long periods for modest lags.
Although advocated by Marsaglia,\cite{SWC} there were no known
published results on statistical tests of the SWC or SWCW generators
prior to the results of Ferrenberg {\it et al.}, so again there was little
support for their claim that these are ``good'' generators.
Recently the shift-with-carry generators were in fact shown to perform
poorly in standard statistical tests.\cite{FINNS}

In this paper the work of Ferrenberg {\it et al.} has been extended by
studies of both the ``good'' generators of that paper, and some ``better''
generators, which are listed below.
In this work there are also more, and in some cases longer, independent runs
for each generator, to obtain better error estimates and to better explore
the effect of different initial seeds.

In a recent review of random number generators,\cite{JAMES} James recommends
3 mixed generators:
\begin{enumerate}
\item   RANECU, L'Ecuyer's mixed LCG combining L(40014,0,2147483563) and \\
	L(40692,0,2147483399).\cite{LECUYER}
\item   RANMAR, Marsaglia's combined LFG F($97,33,-$) and Weyl
	generator.\cite{RANMAR}
\item   RCARRY, a subtract-with-carry generator\cite{SWC}
	based on F($24,10,-$) (this is the same as SWC but with
	a smaller lag).
\end{enumerate}
We also tested the above generators, plus the following:
\begin{enumerate}{\setcounter{enumi}{3}}
\item   RAND, the default 32-bit C and Unix generator
        L($1103515245,12345,2^{31}-1$).
\item   DRAND48, another standard C and Unix generator
	with larger modulus and period, based on
	L({\it 5DEECE66D}$_{16},B_{16},2^{48}$).
\item   RANF, another 48-bit LCG, L({\it 2875A2E7B175}$_{16},0,2^{48}$),
        which is the standard generator used on CRAY and CDC CYBER
	machines.\cite{ANDERSON}
\item   RAN2, which is RANECU augmented by shuffling the order of the
	output values.\cite{NUMREC}
\item   LFGs of different lags, using $+,-,*$ and $\oplus$.
\item   LFGs using $+$ and $\oplus$ with 4 ``taps'',\cite{GOLOMB,ZIFF1,ZIFF2}
	i.e.
        \begin{description}
	\item[\qquad\qquad]
	$X_{i} = X_{i-P} \odot X_{i-Q} \odot X_{i-R} \odot X_{i-S}$,
        \end{description}
	which we will denote by F($P,Q,R,S,\odot$).
\end{enumerate}

We followed Marsaglia and James by initializing each bit of the seed tables
in the LFGs by using a combination LFG and LCG
(see the routines RSTART in
% Ref.~\citelow{RANMAR} and RMARIN in Ref.~\citelow{JAMES}).
Ref.~28 and RMARIN in Ref.~3).
We also tried using RAND to initialize every element of the seed tables,
or every bit of every element in the seed tables, which had little or no
effect on the quality of the LFGs.

For each random number generator, 25 independent simulation runs
with different initial seeds were performed,
on a network of IBM RS/6000, HP Apollo 9000, and DEC 5000 workstations.
Each simulation was between $10^6$ and $5 \! \times \! 10^7$ sweeps of a
$16 \! \times \! 16$ lattice at the critical point of the 2-$d$ Ising
model.\cite{MONTECARLO,FISHER}
The number of random numbers generated per sweep per site varies with the
Monte Carlo algorithm used, with an average of 0.87 for Metropolis, 0.93
for Wolff, and 1.85 for Swendsen-Wang.
For the Metropolis algorithm we chose to visit the sites to be updated in
order, rather than randomly,
to provide a more effective way of probing any regularity or
lattice structure in the sequence of random numbers, especially for the
linear congruential style generators which are known to suffer from
this problem.\cite{MARSAGLIA,KALLE,PLANES,LECUYER,NIEDERREITER,ANDERSON}

Error estimates for each simulation were obtained by standard methods of
binning the data, with a bin size much greater than the autocorrelation
time.\cite{MONTECARLO}
The error in the mean of the 25 combined results was also calculated,
treating them as independent data sets.
Two measures were used to compare the Monte Carlo results with the
exact results:
the deviation $\Delta$ between the mean of the combined results and the
exact value as a multiple of the error in the mean $\sigma$,
and the chi squared per degree of freedom $\chi^2$ for the 25 data
sets.\cite{STATS}
The first test checks for any bias in the average over all runs,
the second checks for discrepancies in the statistical fluctuations
expected between the individual runs.
A generator is judged to have failed the Monte Carlo test if
$\Delta > 3.3 \sigma$, $\chi^2 > 2.0$,  or $\chi^2 < 0.34$,
all of which should occur with probability less than 0.001
for a truly random generator.\cite{STATS}

%%%%%%%%%%%%%%%%%%%%%%%%%%%%%%%%%%%%%%%%%%%%%%%%%%%%%%%%%%%%%%%%%%%
\section{Results}
%%%%%%%%%%%%%%%%%%%%%%%%%%%%%%%%%%%%%%%%%%%%%%%%%%%%%%%%%%%%%%%%%%%
\noindent
The results for $\Delta / \sigma$, the difference between the exact and
simulated values of the energy and specific heat given as a multiple of
the errors in the mean, are presented in Tables \ref{tab1} and \ref{tab2},
along with the values of $\chi^2$. Failure of a test is indicated in bold type.
The generators are grouped into 4 categories, determined by a different
level of precision of the simulations.
Table~\ref{tab1} shows generators which we would classify as bad or
very bad (at least for this type of Monte Carlo application).
The very bad generators failed at least one of the tests with $10^6$
Monte Carlo sweeps per run, with the bad generators failing after
$10^7$ sweeps per run.

Table~\ref{tab2} shows generators which we would classify as good or
very good. The good generators failed one of the tests at a level of
$5 \! \times \! 10^7$ sweeps for the Wolff and Metropolis algorithms,
and $10^7$ sweeps for the SW algorithm (which uses about twice as
many random numbers per sweep).
The very good generators passed all the tests at this level,
which involves generating on the order of
$10^{10}$ random numbers for each of the 25 independent simulations,
or approximately $3 \! \times \! 10^{11}$ random numbers in total.
In contrast, the errors caused by using very bad generators were
generally apparent after using less than $10^9$ random numbers, in
simulations which took only about an hour on a workstation.

\begin{figure}[htbp]
\begin{center}
\begin{picture}(350,275)
\psfig{figure=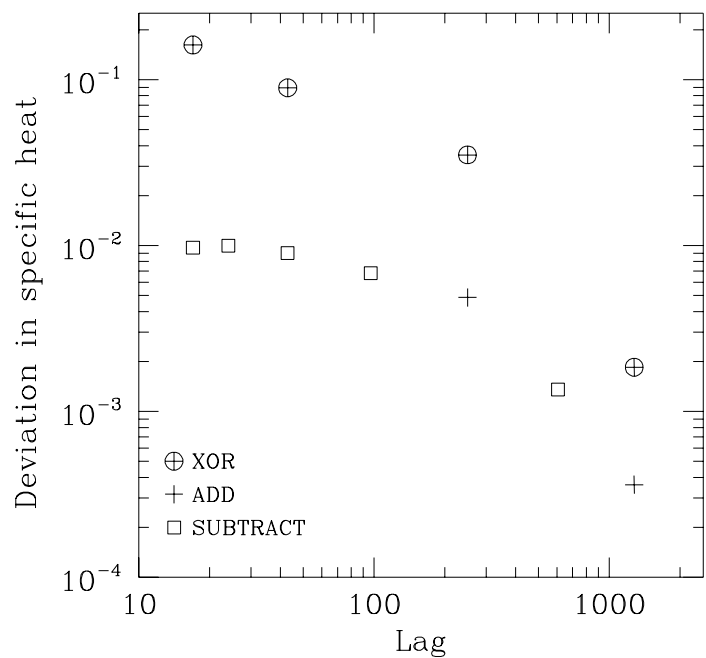,width=350pt,height=350pt}
\end{picture}
\end{center}
\caption{
Relative deviation in the Monte Carlo result for the specific heat of the
2-$d$ Ising model, for the Wolff algorithm using a lagged Fibonacci generator.
Each point denotes a different lag and a different binary operation for the
random number generator.
}
\protect\label{fig1}
\end{figure}

Fig.~\ref{fig1} shows the relative error in the specific heat for the Wolff
algorithm versus the lag of the Fibonacci generator,
for the binary operations addition, subtraction, and XOR.
In all cases the XOR operation was about an order of magnitude worse
than addition and subtraction.
Since in Monte Carlo simulation an order of magnitude decrease in the error
requires 100 times as many iterations, the difference between the quality of
the LFG with different operations is substantial.
Quite large lags of at least 1000 are required to reduce the error
to less than $0.1\%$, however the percentage error for a given lag $P$ goes
roughly as $e^{-P}$, so performance can be greatly improved with
a moderate increase in the lag.
For a lag of 4423 the generators gave correct results for all binary
operations within the errors of the simulations.

Table~\ref{tab3} compares the results for the Wolff algorithm
for various generators based on F($43,22,\odot$), where the binary
operation is XOR, subtraction, subtract-with-carry, and multiplication.
The results of combining this lagged Fibonacci generator
with a Weyl generator (as in SWCW or RANMAR) are also shown.
We can see that the shift register generator using XOR performs very poorly,
with errors of nearly $10\%$ in the specific heat.
Using subtraction performs an order of magnitude better, however
adding a carry bit does not provide any extra improvement.
Mixing in the Weyl generator reduces the errors by nearly another order
of magnitude.
Using multiplication instead of subtraction produces the most dramatic
improvement, for little extra computational cost on modern RISC
workstations.
In Table~\ref{tab4} the standard 2-tap LFG is compared to a 4-tap
version of the same lag, which gives substantially better results,
as was seen by Ziff for self-avoiding random walks.\cite{ZIFF2}

The two 32-bit LCG generators both gave consistent results at the level of
$10^6$ sweeps, for which the number of random numbers required for each
simulation is less than the period of these generators.
Both failed the tests at the level of $10^7$ sweeps, which requires
producing about as many random numbers as the period.
This suggests that the failure is due to the short
period of these generators rather than the lack of randomness.
This is supported in the case of RAND by the fact that some of the $\chi^2$
values in Table~\ref{tab1} are smaller than expected, i.e. the deviations
from the exact value of all the independent runs are {\it too small}.
This is probably due to the fact that each run exhausts the period,
so that different runs are using similar sequences of random numbers and
are therefore correlated to some extent.

The mixed LCG generators RANECU and RAN2 were among the best generators,
although they were also the slowest. This good performance was
rather unfortunate in the case of the RAN2 generator, since
the authors of Numerical Recipes have guaranteed RAN2 to produce
``perfect'' random numbers, with perfect defined as
``we will pay \$1000 to anyone who convinces us otherwise (by finding a
statistical test that RAN2 fails in a non-trivial way, excluding the
ordinary limitations of a machine's floating point
representation).''\cite{NUMREC}

The subtract-with-carry generators RCARRY and SWC were among the worst of the
generators tested, which agrees with the results of
% Refs.~\citelow{LANDAU} and \citelow{FINNS}.
Refs.~11 and 22.
With the notable exception of the version using
multiplication, the lagged Fibonacci generators performed very poorly for
lags under 100 (under 1000 for the case of $\oplus$),
and non-random effects were measurable even for lags of over 1000.
In contrast, standard statistical tests by Marsaglia gave good results for
LFGs using subtraction,
even for lags less than 100 (except for the ``birthday spacings''
test).\cite{MARSAGLIA,RANMAR}
Marsaglia found that LFGs using multiplication performed very well in
statistical tests even for small lags, and this is also true for the Monte
Carlo tests, where multiplication gave by far the best performance for a
given lag.
Generators based on LFGs performed worst for the Wolff algorithm, with some
small lag generators also failing the test with the Metropolis algorithm.
LCGs performed worst on the Metropolis algorithm.

Grassberger\cite{GRASS2} tested F($250,103,\odot$) using Monte
Carlo simulations of random walks, and conjectured that this generator
has large correlations over long times which should only be seen for
Ising model simulations using lattices larger than $16^2$.
We have also done simulations on a $128^2$ lattice to compare the
corresponding errors. The statistical error in the mean energy is
\begin{eqnarray}
\sigma & = & \sqrt {2 * \tau_{int} * variance / sweeps}     \nonumber \\
       & = & \sqrt {2 * \tau_{int} * C_{H} / (V * sweeps)}, \nonumber
\end{eqnarray}
where $\tau_{int}$ is the integrated autocorrelation time,\cite{MONTECARLO,TAU}
$C_{H}$ is the specific heat, and $V$ is the lattice volume.
Since $\tau_{int}$ for the Wolff algorithm has been measured to be
2.6 for $V=16^2$ and 8.2 for $V=128^2$,\cite{POTTS}
and the specific heat increases by 1.69 for the larger lattice,\cite{FISHER}
the statistical error in the mean energy will be approximately the same for
$8.5 \! \times \! 10^4$ sweeps of the $128^2$ lattice as for $10^6$ sweeps
of the $16^2$ lattice, which was indeed found to be the case in our
simulations.
Table~\ref{tab5} shows that the discrepancy in the average energy caused by the
random number generator is actually much smaller for the larger lattice size.
Since for the specific heat the statistical error increases even more rapidly
with increasing lattice volume, smaller lattices seem to be more effective
for testing some random number generators using Monte Carlo simulations of the
Ising model. Of course the inverse result is also true -- some random number
generators will perform better in Monte Carlo simulations on large lattices.

%%%%%%%%%%%%%%%%%%%%%%%%%%%%%%%%%%%%%%%%%%%%%%%%%%%%%%%%%%%%%%%%%%%
\section{Conclusions}
%%%%%%%%%%%%%%%%%%%%%%%%%%%%%%%%%%%%%%%%%%%%%%%%%%%%%%%%%%%%%%%%%%%
\noindent
Lagged Fibonacci generators using the operations of addition, subtraction
or XOR (exclusive OR) can give poor performance, especially for the
Wolff algorithm, unless the lag is very large.
Using addition or subtraction gives substantially better performance than
the shift register generators using XOR.
Using multiplication gives extremely good performance even for small lags.
Adding a carry bit to an LFG using subtraction (the subtract-with-carry
generators) gives no improvement in the performance of these generators,
however adding a simple Weyl generator greatly improves the quality of the LFG.

The multiplicative lagged Fibonacci generator F($P,Q,*$) was one of the best
generators we tested. This generator showed excellent
randomness properties even for very small lags, with only a slightly
greater computational cost than an LCG, or (on modern RISC processors)
an LFG using addition or subtraction.
A multiplicative LFG can be given an arbitrarily large period by simply
increasing the lag. A lag of only 43 gives a period of order $10^{21}$
for 32-bit integer arithmetic, and extremely good randomness properties.
The only drawback of these generators is the lack of a solid theoretical
understanding of their properties.
More theoretical studies and experimental tests should be done on these
generators, since they appear to be very promising candidates for a good
general purpose random number generator.

The 32-bit linear congruential generators perform well up to the point where
their period is exhausted, with RAND seemingly better than CONG. The 48-bit
LCGs such as DRAND48 gave excellent results, and have a large enough period
(of order $10^{14}$) for most current applications. LCGs using even
larger integers, such as L($13^{13},0,2^{59}$), show very good performance in
standard statistical tests,\cite{FINNS,ANDERSON} and have even longer periods.
These longer period LCGs usually require multi-precision arithmetic and are
therefore relatively slow (e.g. DRAND48 is 6 times slower than RAND on a
DECstation 5000), however they should become more popular in the near future,
when 64-bit microprocessors become commonplace.
Apart from an increased period, large $M$ LCGs also have better spectral
(lattice) properties, however the correlations inherent in LCGs are still
present.
Combining a good LCG with another generator, such as an LFG or another
LCG (as with RANECU and RAN2), may further reduce (or even eliminate)
these correlations, however it is possible that this may introduce
other unknown defects. Again, we are hampered by the lack of a good
theoretical understanding of these algorithms.
In general it is probably advisable to stick with a good large $M$ LCG,
which should work perfectly well for most applications.
However it is known that these generators can perform poorly on vector and
parallel computers, where the power-of-2 correlations can be accentuated.
\cite{KALLE,FILK}

Note that by the year 2000 supercomputers will have Teraflop
($10^{12}$ floating point operations per second) performance,
and a Teraflop-year of computation ($3 \! \times \! 10^{19}$ flops)
will become realizable for such problems as Monte Carlo simulation of
lattice QCD and condensed matter physics.\cite{TERAFLOP}
It is therefore likely that large scale Monte Carlo simulations
only ten years from now will exhaust the period (of roughly $10^{18}$)
of 64-bit LCGs or mixed 32-bit LCGs. However a 96-bit or 128-bit LCG, or
a mixed generator made up of two 64-bit LCGs (similar to the RANECU
generator studied here), should have both the randomness properties
and the extremely large period necessary for any application in the
forseeable future.
These multi-precision arithmetic and mixed LCG algorithms are the slowest of
the algorithms tested here, however
it should be noted that the speed of a random number generator is often
irrelevant, since in most applications the amount of time spent generating
the random numbers is insignificant compared to the rest of the calculation.
In most applications the quality of the random numbers is far more important
than the speed with which they are generated.

Mixed lagged Fibonacci generators such as RANMAR have extremely long
periods ($10^{43}$ for RANMAR), however for high precision work the
generator F($97,33,-$) on which RANMAR is based should be
replaced by a longer lag generator with better randomness properties,
such as F($250,103,-$), F($607,273,-$), or F($1279,1033,-$).
The extra memory requirement is negligible for current workstations and
high performance computers, except perhaps for fine grained massively parallel
machines with limited memory per processor.
Mixed generators offer a greatly increased period, and empirical tests
indicate that they can have better randomness properties than the single
generators on which they are based.
The mixed generators were among the best tested here,
however they are not as theoretically well understood as single generators,
so it is possible that unexpected correlations may occur. They should
therefore be used with caution.

Our theoretical understanding of random number generators is quite limited,
and no amount of statistical testing can ever determine the quality of a
generator. It is therefore prudent in any stochastic simulation to use at
least two very different generators (for example, a good large $M$ LCG,
a multiplicative LFG, or a good mixed generator such as RANMAR or RANECU)
and compare the results obtained with each, in order to be confident that
the random number generator is not introducing a bias in the results.

Finally, we should note that it is unfortunate that most of the
poorly performing generators tested here are recommended in many texts
and are available by default to the unwary user on many computer
systems.\cite{KNUTH,PARK}
It should be no more acceptable for a computing environment to have a default
random number generator that is known to be bad, than to have an incorrect
implementation of a standard mathematical function.
Since faster computers and better algorithms are improving the precision of
Monte Carlo and other stochastic simulations at a rapid pace, it is important
to continue to search for better random number generators with very long
periods, and to make more precise and varied tests of these generators.
This is particularly true for high performance computers with vector or
parallel architectures, where methods for generating independent random
numbers in parallel are required.\cite{ANDERSON,NEXT}

%%%%%%%%%%%%%%%%%%%%%%%%%%%%%%%%%%%%%%%%%%%%%%%%%%%%%%%%%%%%%%%%%%%
\section{Acknowledgements}
%%%%%%%%%%%%%%%%%%%%%%%%%%%%%%%%%%%%%%%%%%%%%%%%%%%%%%%%%%%%%%%%%%%
\noindent
I would like to thank John Apostolakis, Barbara Davies, Enzo Marinari,
Alan Sokal and
Robert Ziff for their input, including helpful discussions and suggestions,
reviewing the manuscript, and helping with some of the programs.
The simulations were run on the workstation network of the CASE Center
and the Northeast Parallel Architectures Center (NPAC) at Syracuse University.
Work supported in part by the Center for Research on Parallel Computation
with NSF cooperative agreement No. CCR-9120008,
and by Department of Energy grants
DE-FG03-85ER25009 and DE-AC03-81ER40050.

\newpage

\begin{table}[htbp]
\renewcommand{\arraystretch}{1.2}
\caption{
Results of Monte Carlo simulations of the 2-d Ising model using different
random number generators. The first line for each generator shows
the deviation of the Monte Carlo results from the exact values,
as a multiple of the error in the mean. The second line shows the $\chi^2$
per degree of freedom. Numbers in bold type indicate results
which should occur with a statistical probability of less than 0.001.
This table shows ``bad'' or ``very bad'' generators, grouped as to whether
they failed the test at the level of $10^6$ (very bad) or $10^7$ (bad) sweeps.
}
\vspace*{0.15truein}
\protect\centerline{
\begin{tabular}{|c|c|ccc|ccc|}
\hline %%%%%%%%%%%%%%%%%%%%%%%%%%%%%%%%%%%%%%%%%%%%%%%%%%%%%%%%%%%%%%%%%%
% & &  Energy & &  Specific Heat & & \\
& & &  Energy & &
\multicolumn{3}{c|}{Specific Heat} \\
\hline %%%%%%%%%%%%%%%%%%%%%%%%%%%%%%%%%%%%%%%%%%%%%%%%%%%%%%%%%%%%%%%%%%
Sweeps &
Generator  \quad & \quad SW \quad & Wolff \quad & Metrop
	   \quad & \quad SW \quad & Wolff \quad & Metrop \\
\hline %%%%%%%%%%%%%%%%%%%%%%%%%%%%%%%%%%%%%%%%%%%%%%%%%%%%%%%%%%%%%%%%%%
$10^6$ &
RCARRY
&      ~~0.68  & {\bf ~-9.83} & {\bf -12.21} \quad &
  {\bf ~~7.86} & {\bf ~15.31} & {\bf ~~5.27} \quad \\
&&     ~~1.04  & {\bf ~~7.80} & {\bf ~~3.90} \quad &
  {\bf ~~2.08} & {\bf ~14.83} & {\bf ~~2.35} \quad \\
&
SWC
&      ~~2.00  & {\bf ~-7.66} &      ~~1.18  \quad &
       ~~2.30  & {\bf ~13.49} &      ~~1.13  \quad \\
&&     ~~0.82  & {\bf ~~4.65} &      ~~0.61  \quad &
       ~~1.02  & {\bf ~~9.77} &      ~~1.27  \quad \\
&
F(250,103,$\oplus$)
&      ~-3.13  & {\bf ~32.26} &      ~~0.30  \quad &
       ~-2.33  & {\bf -70.08} &      ~~0.23  \quad \\
&&     ~~0.62  & {\bf ~31.52} &      ~~1.06  \quad &
       ~~1.31  & {\bf 230.47} &      ~~1.15  \quad \\
&
F(250,103,$-$)
&      ~~0.48  & {\bf ~-3.86} &      ~-0.71  \quad &
       ~~1.42  & {\bf ~11.85} &      ~~0.79  \quad \\
&&     ~~1.02  &      ~~0.87  &      ~~0.93  \quad &
       ~~0.92  & {\bf ~~4.06} &      ~~0.92  \quad \\
&
F(250,103,$+$)
&      ~-1.67  &      ~-3.18  &      ~~0.08  \quad &
       ~~1.42  & {\bf ~~9.97} &      ~~0.02  \quad \\
&&     ~~1.37  &      ~~1.23  &      ~~0.58  \quad &
       ~~1.24  & {\bf ~~3.85} &      ~~0.70  \quad \\
\hline %%%%%%%%%%%%%%%%%%%%%%%%%%%%%%%%%%%%%%%%%%%%%%%%%%%%%%%%%%%%%%%%%%
$10^7$ &
RAND
&      ~~1.51  &      ~~0.88  &      ~-0.75  \quad &
       ~-1.46  &      ~-0.07  & {\bf ~-6.61} \quad \\
&&     ~~0.72  & {\bf ~~0.30} & {\bf ~~0.26} \quad &
       ~~1.51  &      ~~0.36  &      ~~1.02  \quad \\
&
CONG
&      ~-0.12  &      ~~0.29  &      ~-1.90  \quad &
       ~-2.88  &      ~-0.80  & {\bf ~~4.92} \quad \\
&&     ~~1.65  &      ~~1.03  & {\bf ~24.64} \quad &
       ~~1.70  & {\bf ~~7.81} & {\bf ~63.56} \quad \\
&
SWCW
&      ~-1.24  &      ~-2.39  &      ~-0.84  \quad &
       ~-0.67  & {\bf ~~4.10} &      ~~0.92  \quad \\
&&     ~~1.41  &      ~~1.16  &      ~~1.72  \quad &
       ~~1.12  &      ~~0.90  &      ~~1.51  \quad \\
&
F(1279,1063,$\oplus$)
&      ~-2.39  & {\bf ~~3.82} & {\bf ~~3.73} \quad &
       ~-2.10  & {\bf -11.78} &      ~-2.51  \quad \\
&&     ~~1.06  &      ~~1.28  &      ~~1.78  \quad &
       ~~0.89  & {\bf ~~5.86} &      ~~1.04  \quad \\
&
F(55,24,16,8,$\oplus$)
&      ~-1.56  & {\bf ~-4.08} &      ~~0.78 \quad &
       ~-3.03  & {\bf ~12.73} &      ~~1.91  \quad \\
&&     ~~1.30  & {\bf ~~4.10} &      ~~1.31  \quad &
       ~~1.57  & {\bf ~14.84} &      ~~1.04  \quad \\
\hline %%%%%%%%%%%%%%%%%%%%%%%%%%%%%%%%%%%%%%%%%%%%%%%%%%%%%%%%%%%%%%%%%%
\end{tabular}}
\protect\label{tab1}
\end{table}

\begin{table}[htbp]
\renewcommand{\arraystretch}{1.2}
\caption{
As for Table~1, except here the number of sweeps is $5 \! \times \! 10^7$
for the Metropolis and Wolff algorithms, and $10^7$ for Swendsen-Wang.
This table shows ``good'' or ``very good'' generators, where
the first (good) group of generators failed some tests at this level,
while the second (very good) group passed all tests.
}
\vspace*{0.15truein}
\protect\centerline{
\begin{tabular}{|c|c|ccc|ccc|}
\hline %%%%%%%%%%%%%%%%%%%%%%%%%%%%%%%%%%%%%%%%%%%%%%%%%%%%%%%%%%%%%%%%%%
% & &  Energy & &  Specific Heat & & \\
& & &  Energy & &
\multicolumn{3}{c|}{Specific Heat} \\
\hline %%%%%%%%%%%%%%%%%%%%%%%%%%%%%%%%%%%%%%%%%%%%%%%%%%%%%%%%%%%%%%%%%%
Sweeps &
Generator  \quad & \quad SW \quad & Wolff \quad & Metrop
	   \quad & \quad SW \quad & Wolff \quad & Metrop \\
\hline %%%%%%%%%%%%%%%%%%%%%%%%%%%%%%%%%%%%%%%%%%%%%%%%%%%%%%%%%%%%%%%%%%
$5 \! \times \! 10^7$ &
RANMAR
&      ~~0.12  &      -0.50  &      ~-0.65  \quad
&      ~~0.75  & {\bf ~5.40} &      ~~0.84  \quad \\
($10^7$ SW)
&&     ~~0.66  &      ~1.01  &      ~~0.94  \quad
&      ~~1.14  &      ~1.19  &      ~~0.91  \quad \\
&
F(1279,1063,$+$)
&      ~~1.38  & {\bf -4.20} &      ~~2.19  \quad
&      ~-0.24  & {\bf ~6.46} &      ~~0.34  \quad \\
&&     ~~0.87  &      ~1.41  &      ~~1.34  \quad
&      ~~0.75  &      ~1.14  &      ~~0.93  \quad \\
&
F(2,1,$*$) $+$ Weyl
&      ~-0.55  &      ~0.79  &      ~-2.45  \quad
&      ~-0.91  &      -0.93  &      ~~0.22  \quad \\
&&     ~~0.88  &      ~1.12  &      ~~0.58  \quad
&      ~~1.19  & {\bf ~2.64} &      ~~1.05  \quad \\
\hline %%%%%%%%%%%%%%%%%%%%%%%%%%%%%%%%%%%%%%%%%%%%%%%%%%%%%%%%%%%%%%%%%%
$5 \! \times \! 10^7$ &
F(4423,1393,$+$)
&  ~~0.82  &  -0.10  &  ~-1.67  \quad
&  ~~1.96  &  ~1.04  &  ~~0.17  \quad \\
($10^7$ SW)
&& ~~0.59  &  ~0.87  &  ~~0.89  \quad
&  ~~1.31  &  ~1.08  &  ~~0.72  \quad \\
&
F(4423,1393,$\oplus$)
&  ~-0.85  &  -1.36  &  ~~1.71  \quad
&  ~~0.53  &  -0.08  &  ~-1.62  \quad \\
&& ~~0.89  &  ~0.87  &  ~~0.72  \quad
&  ~~0.88  &  ~0.97  &  ~~1.14  \quad \\
&
F(5,2,$*$)
&  ~-0.70  &  -2.05  &  ~-0.60  \quad
&  ~-0.23  &  ~2.32  &  ~~0.24  \quad \\
&& ~~1.06  &  ~1.04  &  ~~1.28  \quad
&  ~~1.00  &  ~0.46  &  ~~0.92  \quad \\
&
F(43,22,$*$)
&  ~-0.99  &  -0.52  &  ~-1.47  \quad
&  ~-0.91  &  ~1.21  &  ~~1.23  \quad \\
&& ~~1.09  &  ~1.22  &  ~~0.91  \quad
&  ~~0.73  &  ~1.39  &  ~~0.94  \quad \\
&
F(55,24,16,8,$+$)
&  ~-0.52  &  -0.70  &  ~~1.34  \quad
&  ~~0.63  &  -1.60  &  ~-0.02  \quad \\
&& ~~0.66  &  ~0.88  &  ~~1.54  \quad
&  ~~1.21  &  ~0.92  &  ~~0.83  \quad \\
&
F(218,95,39,11,$\oplus$)
&  ~-0.49  &  ~0.71  &  ~-0.24  \quad
&  ~~0.78  &  -0.75  &  ~~0.00  \quad \\
&& ~~0.81  &  ~1.01  &  ~~0.90  \quad
&  ~~0.43  &  ~1.20  &  ~~1.32  \quad \\
&
RANECU
&  ~~1.29  &  -1.54  &  ~~0.89  \quad
&  ~-0.61  &  ~1.51  &  ~-0.21  \quad \\
&& ~~1.11  &  ~1.44  &  ~~1.14  \quad
&  ~~1.73  &  ~0.79  &  ~~0.76  \quad \\
&
RAN2
&  ~~0.07  &  -2.19  &  ~-2.04  \quad
&  ~-1.51  &  ~1.06  &  ~~2.38  \quad \\
&& ~~1.36  &  ~0.69  &  ~~0.98  \quad
&  ~~0.92  &  ~0.83  &  ~~1.14  \quad \\
&
DRAND48
&  ~~0.10  &  -1.39  &  ~~0.14  \quad
&  ~-0.16  &  ~0.40  &  ~-2.43  \quad \\
&& ~~1.11  &  ~0.65  &  ~~0.61  \quad
&  ~~1.42  &  ~1.56  &  ~~0.56  \quad \\
&
RANF
&  ~~0.37  &  -0.23  &  ~-1.64  \quad
&  ~~0.56  &  ~0.21  &  ~~1.85  \quad \\
&& ~~1.18  &  ~0.70  &  ~~0.88  \quad
&  ~~0.90  &  ~1.00  &  ~~1.12  \quad \\
\hline %%%%%%%%%%%%%%%%%%%%%%%%%%%%%%%%%%%%%%%%%%%%%%%%%%%%%%%%%%%%%%%%%%
\end{tabular}}
\protect\label{tab2}
\end{table}

\begin{table}[htbp]
\renewcommand{\arraystretch}{1.2}
\caption{
Percentage deviation of the Wolff Monte Carlo results from the
exact values for the energy and specific heat of the 2-d Ising model
using different random number generators based on the lagged
Fibonacci generator F(43,22,$\odot$). The binary operations tested
were $-$, $*$, $\oplus$, and subtract-with-carry (SWC).
A Weyl generator was also added to SWC (SWCW) and to F(43,22,$-$) (Weyl).
}
\vspace*{0.15truein}
\protect\centerline{
\begin{tabular}{|c|c|c|}
\hline %%%%%%%%%%%%%%%%%%%%%%%%%%%%%%%%%%%%%%%%%%%%%%%%%%%%%%%%%%%%%%%%%%
\quad Generator \quad  &  \quad Energy \quad  &  Specific Heat  \\
\hline %%%%%%%%%%%%%%%%%%%%%%%%%%%%%%%%%%%%%%%%%%%%%%%%%%%%%%%%%%%%%%%%%%
F(43,22,$\oplus$)  & ~~~0.39~~    & ~~~9.34~ \\
F(43,22,$-$)       & ~~~0.034~    & ~~~0.80~ \\
SWC                & ~~~0.048~    & ~~~0.80~ \\
SWCW               & ~~~0.0039    & ~~~0.057 \\
Weyl               & ~~~0.0039    & ~~~0.058 \\
F(43,22,$*$)       & ~~$<0.002~~$ & ~$<0.02~~~$ \\
\hline %%%%%%%%%%%%%%%%%%%%%%%%%%%%%%%%%%%%%%%%%%%%%%%%%%%%%%%%%%%%%%%%%%
\end{tabular}}
\protect\label{tab3}
\end{table}

\begin{table}[htbp]
\renewcommand{\arraystretch}{1.2}
\caption{
Percentage deviation of the Wolff Monte Carlo results from the exact values
for the energy and specific heat of the 2-d Ising model
using the standard 2-tap lagged Fibonacci generator F(55,24,$\odot$) and
the 4-tap generator F(55,24,16,8,$\odot$).
}
\vspace*{0.15truein}
\protect\centerline{
\begin{tabular}{|c|c|c|}
\hline %%%%%%%%%%%%%%%%%%%%%%%%%%%%%%%%%%%%%%%%%%%%%%%%%%%%%%%%%%%%%%%%%%
\quad Generator \quad  &  \quad Energy \quad  &  Specific Heat  \\
\hline %%%%%%%%%%%%%%%%%%%%%%%%%%%%%%%%%%%%%%%%%%%%%%%%%%%%%%%%%%%%%%%%%%
F(55,24,$\oplus$)       & ~~~0.34~~    & ~~~8.25~ \\
F(55,24,16,8,$\oplus$)  & ~~~0.011~    & ~~~0.29~ \\
F(55,24,$-$)            & ~~~0.028~    & ~~~0.70~ \\
F(55,24,16,8,$+$)       & ~~$<0.002~~~$ & ~~$<0.02~~~$ \\
\hline %%%%%%%%%%%%%%%%%%%%%%%%%%%%%%%%%%%%%%%%%%%%%%%%%%%%%%%%%%%%%%%%%%
\end{tabular}}
\protect\label{tab4}
\end{table}

\begin{table}[htbp]
\renewcommand{\arraystretch}{1.2}
\caption{
Deviation of the Wolff Monte Carlo results from the exact values,
as a multiple of the error in the mean,
using the lagged Fibonacci generators F(250,103,$\oplus$) and
F(250,103,$+$).
The $16^2$ results are for $10^6$ sweeps per run, and the $128^2$
results are for $8.5 \! \times \! 10^4$ sweeps per run.
}
\vspace*{0.15truein}
\protect\centerline{
\begin{tabular}{|c|c|c|c|}
\hline %%%%%%%%%%%%%%%%%%%%%%%%%%%%%%%%%%%%%%%%%%%%%%%%%%%%%%%%%%%%%%%%%%
\quad Generator \quad  & Lattice Size & \quad Energy \quad  &  Specific Heat \\
\hline %%%%%%%%%%%%%%%%%%%%%%%%%%%%%%%%%%%%%%%%%%%%%%%%%%%%%%%%%%%%%%%%%%
F(250,103,$\oplus$)    &  $16^2$   &  32.26  &  -70.08  \\
F(250,103,$\oplus$)    &  $128^2$  &  ~3.26  &  ~-9.31  \\
F(250,103,$+$)         &  $16^2$   &  -3.18  &  ~~9.97  \\
F(250,103,$+$)         &  $128^2$  &  -1.33  &  ~-0.11  \\
\hline %%%%%%%%%%%%%%%%%%%%%%%%%%%%%%%%%%%%%%%%%%%%%%%%%%%%%%%%%%%%%%%%%%
\end{tabular}}
\protect\label{tab5}
\end{table}

\newpage

\end{document}